# Object-Based Audio Rendering

Philip Jackson[1], Filippo Fazi[2], Frank Melchior[3], Trevor Cox[4], Adrian Hilton[1],
Chris Pike[3], Jon Francombe[5], Andreas Franck[2], Philip Coleman[1], Dylan Menzies-Gow[2],
James Woodcock[4], Yan Tang[4], Qingju Liu[1], Rick Hughes[4], Marcos Simon Galvez[2],
Teo de Campos[1], Hansung Kim[1], Hanne Stenzel[1]

24 August 2017

[1] Centre for Vision, Speech and Signal Processing (CVSSP), University of Surrey, UK
[2] Institute of Sound & Vibration Research (ISVR), University of Southampton, UK
[3] Audio Research, BBC R&D, UK
[4] Acoustics Research Centre (ARC), University of Salford, UK
[5] Institute of Sound Recording (IoSR), University of Surrey, UK

## Overview

This document provides a transcript of GB Patent Application No: GB1609316.3, which
was filed in the UK by the University of Surrey on 23 May 2016. It describes an
intelligent system for customising, personalising and perceptually monitoring the
rendering of an object-based audio stream for an arbitrary connected system of
loudspeakers to optimize the listening experience as the producer intended.

## Acknowledgements

The development of the concepts and implementation were supported by the EPSRC
Programme Grant S3A: Future Spatial Audio for an Immersive Listener Experience at
Home (EP/L000539/1). The authors would like to thank Chelsea Brain and Rob Yates
of the University of Surrey and Rob Cork of Venner Shipley LLP for their assistance in
the preparation of this document.

## Object-Based Audio Rendering

### Technical Field

The present invention relates to object-based audio rendering.

### Background

Systems and methods for reproducing audio can broadly be categorised as either channel-based or object-based. In channel-based audio, a soundtrack is created by recording a separate audio track (channel) for each speaker. Common speaker arrangements for channel-based surround sound systems are 5.1 and 7.1, which utilise five and seven surround channels respectively, and one low-frequency channel. A major drawback of channel-based audio is that each soundtrack must be created for a specific speaker configuration, hence the development of industry-standard configurations such as 2.1 (stereo), 5.1 and 7.1.

Object-based audio addresses this drawback by representing a sound scene as multiple separate audio objects, each of which comprises one or more audio signals and associated metadata. Each audio object is associated with metadata that defines a location and trajectory of that object in the scene. Object-based audio rendering involves the rendering of audio objects into loudspeaker signals to reproduce the authored sound scene. As well as specifying the location and movement of an object, the metadata can also define the type of object and the class of renderer that should be used to render the object. For example, an object may be identified as being a diffuse object or a point source object. Typically, object-based renderers use the positional metadata with a rendering algorithm that is specific to the particular object type to pan each object over a wide variety of conformal loudspeaker arrangements, based on knowledge of the loudspeaker directions from the predefined 'sweet spot' listening position.

Object-based renderers provide greater flexibility than channel-based audio systems, insofar as they can cope with different numbers of speakers. In practice, however, existing object-based methods can only cope with a very limited degree of irregularity, and often require a static sweet spot, gain and phase compensation. Existing object-based methods can also suffer from a small sweet spot. There is therefore a need in the art for improved object-based rendering apparatuses and methods.

The invention is made in this context.

**Summary of the Invention**

According to a first aspect of the present invention, there is provided apparatus for performing object-based audio rendering on a plurality of audio objects which define a sound scene, each audio object comprising at least one audio signal and associated metadata, the apparatus comprising: a plurality of renderers each capable of rendering one or more of the audio objects to output rendered audio data; and object adapting means for adapting one or more of the plurality of audio objects for a current reproduction scenario, the object adapting means being configured to send the adapted one or more audio objects to one or more of the plurality of renderers.

In some embodiments according to the first aspect, the object adapting means comprises: a scene adapter configured to adapt the sound scene for the current reproduction scenario by adapting an audio signal and/or metadata of one or more of the audio objects; and an object refiner configured to receive the plurality of audio objects, including any audio objects adapted by the scene adapter, further refine individual ones of the audio objects by adapting an audio signal and/or metadata of said individual audio objects, and send the plurality of audio objects to one or more selected renderers among the plurality of renderers.

The object adapting means may further comprise a contextual information generator configured to determine one or more parameters relating to the current reproduction scenario, and to control the scene adapter and/or the object refiner based on predefined rules for optimising the rendered audio data according to the determined one or more parameters.

In some embodiments according to the first aspect, the metadata includes editorial information defining a constraint within which an audio object may be adapted, and the object adapting means is configured to adapt one of the plurality of audio objects within the constraint defined by the metadata. For example, the editorial information may define a priority order of a plurality of properties of one or more audio objects within the sound scene, and the object adapting means is configured to preferentially modify a property with a lower priority instead of modifying a property with a higher priority, when adapting the one or more audio objects.

In some embodiments according to the first aspect, the associated metadata for each of the plurality of audio objects defines an object type, and the object adapting means is configured to be capable of sending different audio objects of the same type to different renderers.

In some embodiments according to the first aspect, the object adapting means is configured to adapt one or more of the audio objects according to a user preference setting. The user preference setting may, for example, include one or more of: an intelligibility setting defining a preferred intelligibility level for a dialogue audio object; an envelopment setting; and a language setting defining a preferred language.

In some embodiments according to the first aspect, the apparatus further comprises an input configured to receive an audio signal from at least one microphone arranged to detect sound generated by the one or more speakers, wherein the scene adapter is configured to increase intelligibility of a dialogue audio object in response to an increase in a noise level detected by the one or more microphones. In some embodiments according to the first aspect, the scene adapter may be configured to adapt a dialogue audio object so as to increase the intelligibility in response to a relative increase in the level of one or more other audio objects in the scene.

In some embodiments according to the first aspect, the object adapting means is configured to determine the current reproduction scenario based on one or more environment parameters relating to a physical environment in which the rendered audio data is to be reproduced.

In some embodiments according to the first aspect, the object adapting means is configured to determine the current reproduction scenario based on one or more speaker parameters relating to capabilities of one or more speakers through which the rendered audio data is to be reproduced. In such embodiments, the apparatus may further comprise: a device manager configured to receive information identifying audio output capabilities of one or more connected external devices, and set the one or more speaker parameters according to the received information; and a speaker interface configured to send the rendered audio data to the one or more connected external devices for reproduction. For example, the one or more speaker parameters may include one or more of: speaker number information defining a number of speakers presently connected to the speaker interface; speaker location and/or orientation information defining a location and/or orientation of at least one of the one or more speakers; bandwidth information defining a bandwidth of a connection between the speaker interface and one of the one or more speakers; frequency response information defining a frequency response of at least one of the one or more speakers; and latency information defining a response time of at least one of the one or more speakers.

In some embodiments according to the first aspect, the object adapting means is configured to determine the current reproduction scenario based on an audience location parameter defining a location and/or orientation of one or more individuals within a space in which the rendered audio data is to be reproduced.

According to a second aspect of the present invention, there is provided an object-based audio rendering method comprising: receiving a plurality of audio objects which define a sound scene, each audio object comprising at least one audio signal and associated metadata; adapting one or more of the plurality of audio objects for a current reproduction scenario; and sending the adapted one or more audio objects to one or more of a plurality of renderers to be converted into rendered audio data.

In some embodiments according to the second aspect, adapting the one or more audio objects comprises: adapting the sound scene for the current reproduction scenario by adapting an audio signal and/or metadata of one or more of the audio objects; further refine individual ones of the audio objects by adapting an audio signal and/or metadata of said individual audio objects; and sending the plurality of audio objects to one or more selected renderers among the plurality of renderers.

According to the third aspect of the present invention, there is provided a computer-readable storage medium arranged to store computer program instructions which, when executed, perform a method according to the second embodiment.

**Brief Description of the Drawings**

Embodiments of the present invention will now be described, by way of example only, with reference to the accompanying drawings, in which:

Figure 1 illustrates apparatus for object-based audio rendering, according to an embodiment of the present invention;

Figure 2 illustrates the controller of Fig. 1, according to an embodiment of the present invention;

Figure 3 illustrates audio objects of the same type being routed to different renderers, according to an embodiment of the present invention;

Figure 4 is a flowchart showing an object-based sound rendering method, according to an embodiment of the present invention; and

Figure 5 schematically illustrates a metadata structure for providing information about a current reproduction scenario, according to an embodiment of the present invention.

## Detailed Description

Referring now to Fig. 1, apparatus for object-based audio rendering is illustrated according to an embodiment of the present invention. The apparatus 100 is capable of adapting audio objects for a current reproduction scenario, in order to provide optimised output. The object adaption may, for example, be driven by one or more metrics that provide an estimation of how a listener perceives the reproduced sound scene. New values of the one or more metrics can be measured or estimated at regular intervals, for example every few seconds, to allow the adaption to change with a dynamic scene.

The audio reproduction scenario may, for example, be defined in terms of the acoustic properties of the environment 120 in which the rendered audio data is to be reproduced, characteristics of a particular listener or group of listeners for whom the rendered audio data is intended, physical capabilities of speakers or other devices which will be used to reproduce the rendered audio data, and/or the audio content that is to be reproduced. The apparatus 100 may adapt the audio objects based on knowledge of various factors which influence the physical reproduction of the rendered audio data. For example, the rendering process can be adapted according to the requirements of a specific individual, such as a hearing impairment, and/or according to characteristics of the physical environment in which the audio is to be reproduced, such as the actual loudspeaker positions or the reverberation time of the room.

The apparatus 100 is configured to receive a plurality of audio objects 110, each of which comprises at least one audio signal 111 and associated metadata 112. The apparatus 100 is further configured to render the objects 110 into loudspeaker signals comprising rendered audio data. The rendered audio data is outputted to various speakers distributed in the environment 120 in which the audio is to be reproduced. As will become apparent from the following description, embodiments of the present invention can utilise arbitrary configurations of loudspeakers when rendering audio, and can adapt the audio output depending on the particular capabilities of the current speaker configuration.

As described above, in known object-based audio rendering systems the object metadata defines parameters such as the object type, location and trajectory. In some embodiments of the present invention, the metadata 112 for an audio object 110 may also include editorial information defined by the producer. Examples of editorial information that can be defined in embodiments of the present invention include metadata tags to identify a particular object as being a dialogue or background object, and metadata tags to identify the type of environment in which the sound scene is set, such as an expansive outdoor environment or a confined interior environment. Further examples of editorial information include tolerances within which the apparatus 100 may adapt the audio objects, and perceptual priorities which define an order of priority in which the apparatus 100 should attempt to optimise properties such as intelligibility, position, velocity, locatedness, scale, and envelopment of a particular audio object. For example, when intelligibility is set as a high perceptual priority for an audio object, the apparatus 100 may prioritise the intelligibility of that object when determining how to adapt and/or render the object. Such editorial information gives the producer greater control over the narrative function and perceptual properties of each audio object than is possible with conventional metadata schemes.

When the metadata for an audio object defines tolerances within which the apparatus may adapt that audio object, it is possible for a conflict to arise between the tolerances defined by the metadata and other input variables, such as user preferences, which may indicate that the object should be adapted beyond the range of tolerances defined in the metadata. In such scenarios, the apparatus is preferably configured to prioritise the tolerances defined by the metadata, such that the object is only adapted within the defined tolerances.

In the embodiment shown in Fig. 1, the environment 120 in which the audio is to be reproduced includes a plurality of devices which are capable of reproducing sound. The plurality of devices in the present embodiment comprises a display unit 121, for example a digital television with integrated stereo speakers, a plurality of standalone loudspeakers 131, 132, 133, 134, and a plurality of multipurpose devices which include integrated speakers, such as a desktop computer 141, tablet computer 142, laptop computer 143, and smartphone 144. In general, the apparatus 100 may be configured to communicate with output devices via any suitable wired or wireless interfaces. It will be appreciated that the types of device illustrated in Fig. 1 are merely examples, and other types of device may be utilised in other embodiments of the present invention. For example, in some embodiments the apparatus 100 can send audio signals to wearable devices such as headphones or electronic hearing aids, allowing the audio output to be directed to a specific individual.

As shown in Fig. 1, the apparatus 100 of the present embodiment comprises a scene adapter 101 configured to receive the audio objects 110, an object refiner 102 configured to receive adapted objects from the scene adapter 101 and further refine individual objects prior to rendering, a bank of renderers 103 configured to render refined objects received from the object refiner 102, and a speaker interface 104 configured to send the rendered audio signals to the plurality of output devices 121, 131, 132, 133, 134, 141, 142, 143, 144. The apparatus 100 further comprises a context unit 105 which provides contextual information to the scene adapter 101 and object refiner 102 for controlling how the objects 110 are adapted and refined. The contextual information is indicative of the current reproduction scenario, and enables the scene adapter 101 and object refiner 102 to optimise the sound scene and objects 110 for the current reproduction scenario.

Together, the plurality of objects 110 define a sound scene. The scene adapter 101 determines how to modify the sound scene based on contextual information provided by the context unit 105, which conveys information to the scene adapter 101 about the current reproduction scenario. The scene adapter 101 is configured to adapt the audio signals 111 and/or the associated metadata 112 of at least one of the audio objects 110, and may also adapt the scene structure as a whole, for example by regrouping, refactoring or pruning objects. In addition to taking into account contextual information relating to the current reproduction scenario, in some embodiments the scene adapter can also take into account properties of other ones of the objects 110, for example the relative locations of objects, when adapting the scene.

As an example, if the contextual information indicates that a listener has a hearing impairment or that there is a high level of background noise in the environment 120, the scene adapter 101 can adapt the metadata 112 in order to change the positions of objects within the sound scene, or adjust their relative levels, so as to increase the intelligibility of a dialogue audio object. Instead of or in addition to adapting the metadata 112, the scene adapter could adapt the audio signal(s) of a dialogue object by filtering or otherwise processing the signal.

In effect, the scene adapter 101 is able to re-mix the sound scene so as to optimise the audio output for a particular listener, environment, and/or speaker configuration. The scene adapter 101 operates in the object domain, that is, before the objects 110 are processed in the bank of renderers 103. The scene adapter 101 then outputs any objects that have been adapted to the object refiner 102, along with any objects that remain unaltered.

The object router 102 receives the adapted objects from the scene adapter 101, and further refines individual ones of the audio objects 110 before they are sent to the bank of renderers 103. The renderer bank 103 converts the sound scene into speaker signals, and comprises a plurality of rendering algorithms each of which is capable of rendering an audio object. The renderers within the bank 103 may be grouped into various renderer classes, including but not limited to: gain-delay renderers; filter-based renderers; decorrelating renderers; binaural renderers; and multi-zone renderers. Examples of gain-delay renderers include an AP1 renderer (nearest l/s), an AP3 renderer (vector-base amplitude panning, VBAP), an ambisonic mode matching (MM) renderer, and a wavefield synthesis (WFS) renderer. Examples of filter-based renderers include a single zone pressure matching (PM) renderer and a planarity panning (PP) renderer. Examples of decorrelating renderers include a diffuse renderer and a directional audio coding renderer (DirAC). Examples of multi-zone renderers include an acoustic contrast control renderer (ACC), a planarity control renderer (PC), and a multi-zone pressure matching renderer (PM). It will be understood that these are merely a few examples of types of renderers that can be used in embodiments of the present invention, and should not be construed as limiting.

In the present embodiment, the object refiner 102 is configured to select one or more suitable rendering algorithms for converting the adapted audio objects 110 into speaker signals, according to contextual information provided by the context unit 105. In many cases a single rendering algorithm will be selected, but in some instances a plurality of renderers may be selected for a single object. For example, during a period of time in which the system cross fades from one renderer to another, two renderers are used simultaneously during the transition.

The information provided to the scene adapter 101 by the context unit 105 may be referred to as high-level contextual information, that is to say, contextual information that is relevant to the scene as a whole.  The information provided to the object refiner 102 by the context unit 105 may be referred to as low-level contextual information, that is to say, contextual information that is relevant to the rendering of individual ones of the audio objects 110.  The object refiner 102 can utilise the low-level contextual information in order to apply further refinement on an object-by-object basis, whereas the scene adapter 101 operates on the scene as a whole.  The object refiner 102 can further be configured to select the optimum renderer for each object, taking into account the current reproduction scenario.

In the present embodiment, as described above, audio objects are adapted for the current reproduction scenario in a two-stage process, whereby high-level scene adaptation is applied first followed by low-level refinement of individual objects.  In other embodiments, however, a single adaptation stage may be performed with the adapted objects being sent directly to the bank of renderers 103.

After the objects 110 have been rendered, the rendered audio data is sent to the plurality of output devices 121, 131, 132, 133, 134, 141, 142, 143, 144 via the speaker interface 104.  Depending on the requirements of the connected devices, the speaker interface 104 may be configured to output the rendered audio data as digital data or as analogue audio signals.  For example, the speaker interface 104 may include a digital-to-analogue converter (DAC) to convert the rendered audio data into an analogue speaker signal.  Alternatively, when an output device is capable of receiving digital data, the rendered audio data may be transmitted to the output device without being converted into an analogue signal.  For example, the rendered audio data may be transmitted directly to the desktop computer 141, tablet computer 142, laptop computer 143, and smartphone 144 in the embodiment shown in Fig. 1.

In some embodiments the apparatus 100 may further include a device manager 106 for connecting to and managing various output devices in the environment 120.  The device manager 106 can also synchronise the connected devices to ensure that the audio output from all devices in the environment 120 is synchronised.  The device manager 106 can search for, and connect to, various types of devices through the speaker interface 104.  Examples of devices that can be used as output devices in embodiments of the present invention include, but are not limited to, additional displays (e.g. TVs) other than the display unit 121 on which content is currently being displayed, multimedia equipment such as optical disk players, computer monitors, computer games consoles and controllers, internet-enabled radios, wireless loudspeakers, portable media players, hearing aid or headphones, earphones, and headsets with integrated audio such as augmented reality (AR) and virtual reality (VR) headsets. The connected devices will primarily include audio-enabled devices capable of reproducing audio, but can also include devices with other output capabilities such as haptic feedback mechanisms.

The device manager 106 can also connect to devices which include sensors for detecting physical properties of the output environment 120 in which the rendered audio data is to be reproduced. The context unit 105 can receive signals from the sensors in various connected devices in order to learn more about the environment 120, and define the current reproduction scenario accordingly. For example, the context unit 105 may receive signals from such sensors as microphones, gyrometers, accelerometers, still image or video cameras, infrared video, ultrasound, heart rate monitors or other medical instruments. Such sensors can provide feedback about the current state of the environment and/or individual listeners. Furthermore, in some embodiments the control unit 105 may receive feedforward signals containing information about a future state of the environment 120. For example, when the apparatus 100 is implemented in an in-car entertainment (ICE) system, the context unit 105 may receive a feedforward signal from the car management system to indicate that the engine revs are about to be increased or decreased, allowing the context unit 105 to anticipate a change in background noise level.

The connected output devices may also return information specifying their output capabilities to the device manager 106, which can store the information for subsequent use when optimising the rendered audio data. By being able to output rendered audio data to various different types of device, the apparatus 100 can provide a more immersive user experience. For example, when the sound scene includes an audio object for a ringing telephone, the rendered audio data for that audio object can be sent to a smartphone 144.

In selecting the renderer for a particular object 110, the object refiner 102 can be configured to send different objects of the same type to different renderers. An example in which an object refiner 302 sends different objects of the same type to different classes of renderer is shown in Fig. 3, in which solid circles represent objects of one type and solid squares represent objects of another type, and the open circle, triangle and square represent different classes of renderer among a bank of renderers 303. For example, the rendering algorithm for a particular audio object may be selected according to scene properties such as the 3D position, 3D extent, and locatedness or diffuseness of a particular object, and/or according to user preferences. For example, the user preferences may indicate whether the listener has a hearing impairment.

As an example, the object refiner 302 may decide to render two 'dialogue' type objects using different renderers when one of the objects corresponds to a character in the foreground of the scene, and the other object corresponds to a character in the background of the scene. For a narrator, the object refiner 302 could decide to use a renderer that is configured to output a loudspeaker signal to a speaker that is close to the listener, such as a wireless loudspeaker or an integrated speaker within a second-screen device such as a mobile phone or tablet. For an actor on-screen, the object refiner 302 could decide to use a VBAP renderer to pan to the location of the actor in the scene. For background speech, for example in a cocktail party scene, the object refiner 302 could decide to use yet another rendering method such as ambisonic rendering, or a rendering algorithm which is configured to distribute voices over multiple discrete loudspeakers.

By configuring an object refiner 302 to be capable of selecting different rendering algorithms for different audio objects of the same type, embodiment of the invention can be free to select the optimum renderer for each audio object in the scene. As illustrated by the examples above, in some embodiments the decision of rendering algorithm for an object may take into account the specific loudspeakers with which a particular renderer is configured to operate, to provide a more immersive listening experience. In contrast, known object-based audio rendering methods only take into account the object type when selecting a suitable renderer.

Furthermore, in some embodiments the apparatus may be configured to select the optimum renderer for each of the received audio objects 110 based on contextual information defining the current reproduction scenario, without applying any adaptation before rending. In such embodiments the scene adapter 101 may be omitted, and the object refiner 102 may be referred to as an object router which operates to route each object to the selected optimum renderer. For example, as described above, the object router may select an optimum renderer by taking into account contextual information such as the configuration and capabilities of connected loudspeakers, and/or other information about the listener and the environment.

The context unit 105 of Fig. 1 is illustrated in more detail in Fig. 2. In the present embodiment, the context unit 105 comprises a listener information acquiring unit 201, a system configuration information acquiring unit 202, a monitoring unit 203, and a processor 204. The listener information acquiring unit 201, system configuration information acquiring unit 202, and monitoring unit 203 together acquire information that can be used to define the current reproduction scenario. In general, a reproduction scenario can be defined in terms of any factors that will influence the listener's perception of the rendered audio data. The listener information acquiring unit 201, system configuration information acquiring unit 202 pass on the acquired information to the contextual information generator 204.

In the present embodiment, the contextual information generator 204 is configured to provide high-level and low-level contextual information respectively to the scene adapter 101 and the object refiner 102 as described above. The scene adapter 101 and the object refiner 102 can apply predefined rules when deciding how to optimise the rendered audio data for the current reproduction scenario, to ensure that the rendered audio is perceived correctly by the listener. In this way, embodiments of the present invention can ensure that the rendered audio is perceived as intended by the producer, taking into account user preferences such as a preferred language setting or a hearing impairment setting.

The rules applied by the scene adjuster 101 and/or the object refiner 102 can be defined in advance when configuring the apparatus 100, as in the present embodiment. For example, the predefined rules can be configured to replicate the knowledge and judgements as demonstrated by an expert producer or group of producers. In other embodiments the contextual information generator 204 may be configured to operate without predefined rules, for example, the scene adapter 101 and the object refiner 102 may apply machine learning to adapt according to the listener's behaviour. Examples of simple rules that can be applied to determine the rendering instructions to be sent to the scene adapter 101 and/or the object refiner 102 include ducking, attenuating, displacing or decorrelating background music whilst a character in the sound scene is speaking.  This could include a crossfade from one renderer (e.g. VBAP) to another (e.g. Ambisonics).  Another example of a predefined rule could be to associate specific types of content with preferences for certain rendering algorithms, for example, a preference to render speech by a discrete loudspeaker, to render ambience by Ambisonic panning over the backdrop loudspeakers, or to render non-diegetic music by pressure matching for moderately large systems.

In addition to user preferences such as a preferred language setting or a hearing impairment setting, examples of other types of information that can be acquired by the listener information acquiring unit 201 include audience parameters such as the number of listeners, their positions and/or orientations within the audio environment 120, their hearing characteristics, focus of attention, and aspects of their behaviour. The number of listeners and their positions can be determined in various ways, for example by analysing acoustic or visual signals recorded from the environment 120, or by providing a user interface through which a user can input the locations of any listeners.  As an example, a facial recognition algorithm could be applied to an image captured by a webcam in order to identify the number and location of individual listeners within the environment 120.  By acquiring information about the number of listeners and their locations, the apparatus 100 can optimise the rendered audio data for each user.  Different ones of the users may have set different preferences, and personalised audio output can be optimised in accordance with that particular user's preferences.  The personalised audio output could be delivered by sending the corresponding rendered audio data for a user to the speakers closest to that user, and/or by using noise cancelling techniques to create a cancellation zone around other listeners.  In some embodiments the listener information can include information obtained from social media or other sources such as health records, such as the listener's age or other demographic information.

For example, when reproducing audio from TV coverage of a football match, one user may have a preference for one team and another user may have a preference for the other team.  The preference may be manually selected or may be determined automatically, for example by analysing social media accounts for each user to determine their favourite team, or by analysing user behaviour during the match.  In this scenario, the sound scene may include different audio objects corresponding to crowd noise or commentary for the different teams.  The scene adapter 101 can be controlled to increase the volume level of the crowd noise or commentary for the preferred team in the personalised audio output for each listener, providing a personalised experience.

Examples of types of information that can be acquired by the system configuration information acquiring unit 202 include information about the speaker properties, such as the position, connection bandwidth, frequency response, latency, directivity, sensitivity, distortion, and power limits. The information about speaker properties may be acquired automatically from a device manager 106, or may be input by a user through a suitable user interface, or may be obtained from an external database. As a further alternative, in some embodiments the apparatus 100 may include a suitable mechanism for detecting the location and/or properties of the speakers, or may determine the location and/or properties based on signals from sensors within the environment 120 (e.g. acoustic, RF, ultrasonic, or laser sighting sensors). For example, an image may be captured of the environment, and a pattern recognition algorithm may be used to detect the location of speakers within the image. Other properties such as the frequency response and distortion may be determined by sending a known signal to each speaker and measuring the output via one or more microphones 151.

The information acquired by the system configuration information acquiring unit 202 may be particularly advantageous in a system of diverse audio-enabled devices as shown in Fig. 1, since the speaker properties may vary widely from device to device. For example, the object refiner 102 can take into account the availability of mid-range and low-frequency loudspeaker angles with respect to the sweet spot or a particular listener and select an appropriate rendering algorithm for a certain object. Furthermore, in some embodiments the apparatus 100 may further comprise means for identifying a subset of one or more of the plurality of connected output devices 121, 131, 132, 133, 134, 141, 142, 143, 144 which satisfies the arrangement requirements of a given rendering method, to determine which of the output devices will be available for reproducing audio data rendered by the corresponding rendering algorithm in the bank of renderers 103.

Instead of or in addition to information about speaker properties, the system configuration information acquiring unit 202 can also be configured to acquire information about the physical environment in which the rendered audio data will be reproduced. Again, such information may be acquired automatically or inputted by a user, depending on the embodiment. For example, the information about the physical environment may include a size of the environment, acoustic properties of materials in the environment, and/or a location of target artefacts such as phones, TVs, windows and doors. The target artefacts may be acoustically illuminated via a 1-NN loudspeaker selection, backdrop panning (e.g., VBAP), or sound reflection to make it appear to the listener as though a particular audio object is coming from the location of the target artefact, providing a more immersive experience. The information about the physical environment can be referred to as environmental information, and may be used when selecting the renderer for a particular audio object, to avoid unpleasant room reflections for example when driving a sound bar.

As shown in Fig. 1, the context unit 105 can be configured to receive a microphone signal from a microphone 151 that is disposed to capture an audio signal from the environment 120. The microphone signal can be analysed by the monitoring unit 203 to acquire feedback information for use by the contextual information generator 204. In some embodiments, the monitoring unit 203 may be configured to receive signals from a plurality of microphones distributed in the environment 120, for example near-listener spot microphones, loudspeaker-mounted microphones or binaurally-recorded microphones.

Examples of types of information that can be acquired by the monitoring unit 203 from the microphone signal include perceptual attributes such as loudness, timbral quality, envelopment, annoyance, distraction, spatial audio quality, and speech intelligibility. This can be referred to as perceptual monitoring. The context unit 105 can control the object refiner 102 to adapt the choice of rendering algorithm for a particular audio object in response to perceptually-motivated monitoring of the virtual or physical sound field that the listener experiences, and/or local noise conditions, as determined from the microphone signal.

Figure 5 illustrates examples of types of metadata that can be used in embodiments of the present invention. The metadata includes various parameters, which may include metrics that provide an estimation of how a listener perceives the reproduced sound scene. For clarity, in the example shown in Fig. 5 the parameters are grouped into categories according to the type of property to which each parameter relates. The metadata in Fig. 5 includes object metadata that defines properties of an audio object. The object metadata can be subdivided into basic object metadata 501 and advanced object metadata 502. The basic object metadata 501 includes baseline object metadata, which for example may define object properties such as the object type; ID; channels; group; priority; and level. Other object properties that can be defined in the basic object metadata 501 include: point information (e.g. position); plane information (e.g. direction; reference distance); diffuse properties; HOA properties (e.g. rotation; warping); diffuse point information; and extended properties (e.g. extent). Examples of properties that can be defined in the advanced object metadata 502 include: object category; near room feature; importance to narrative; emotional content; onscreen/offscreen; interactivity restriction; object quality; preferred rendering method; target device; bandwidth; signal statistics; and language.

As shown in Fig. 5, in the present embodiment the metadata further comprises additional categories of metadata that is generated by the context unit 105, divided into. The reproduction environment metadata 503, scene metadata 504 and listener metadata 505 may, for example, be generated based on automatic analysis of the reproduction environment, or based on user input.

As shown in Fig. 5, the reproduction environment metadata 503 may be sub-divided into reproduction metadata, loudspeaker metadata, surface metadata and room feature metadata. The reproduction metadata can include such information as: loudspeakers; surfaces; physical objects in the reproduction room; layout limitations; and soundfield-based perceptual data. The loudspeaker metadata can include such information as: loudspeaker ID; position; renderer assignment; quality information; and latency. The surface metadata can include such information as the position and material of one or more surfaces in the reproduction environment 120. The room feature metadata can include such information as a description and position of one or more features in the reproduction environment 120. The scene metadata 504 may include information such as: target envelopment; target intelligibility; and object-based perceptual data. The listener metadata 505 can include information such as: listener position; desired envelopment; accessibility requirements; and preferred language.

In the embodiment shown in Fig. 1, elements of the metadata shown in Fig. 5 can be passed to the scene adapter 101 and the object refiner 102 as high-level and low-level contextual information, respectively. It will be understood that the metadata structure shown in Fig. 5 is merely one example, and other types of parameters may be used in other embodiments of the invention.

In some embodiments of the present invention, the rules applied by the scene adapter 101 and/or object refiner 102 can link the perceptual monitoring with various semantic targets for the desired sound field, including but not limited to accurate positioning of sound objects, the degree of envelopment, intimacy or spaciousness of a scene, the timbral quality, or the intelligibility of speech content over and above the background scene and local noise (e.g., dishwasher or road noise). The rendering algorithm selected by the object refiner 102 can be determined in accordance with these targets, which may be evaluated single-endedly or double-endedly using real and/or virtual signals based on simulations from various stages prior to physical sound reproduction. Here, 'single-endedly' refers to embodiments in which only the output signal is available, and 'double-endedly' refers to embodiments in which both the input and output signals are available. Single-ended perceptual evaluation is based only on the signals arriving at the listener, whereas double-ended evaluation refers to comparison against some reference. For example, an absolute loudness measurement can be made in a single-ended way with a calibrated microphone signal. A double-ended comparison would be required to assess the effect of processing artefacts, by comparing the received signal to the original input signal. Relevant perceptual models for the simulations could include basic audio/speech quality, spatial audio quality, distraction due to an interfering signal, intelligibility or general preference.

As described above, the scene adapter 101 and object refiner 102 can decide how to adapt the scene and individual objects according to predefined rules, based on the information that has been acquired regarding the current reproduction scenario. In any given scenario, both the scene adapter 101 and the object refiner 102 may be used to adapt the audio objects, or only one of the scene adapter 101 and the object refiner 102 may be needed to achieve the necessary optimisation. For example, in certain scenarios the rendered audio data may be optimised by controlling the object refiner 102 to select different renderers, without adapting the objects 110 first. In this scenario, the scene adapter 101 can be disabled such that the objects 110 are passed unaltered to the object refiner 102. Conversely, in another scenario scene adjustment may be performed without subsequent refinement of individual objects, in which case the object refiner 102 may be disabled or omitted entirely. In general the scene adapter, object refiner and context unit can be referred to collectively as an object adapting unit or object adapting means for adapting one or more of the plurality of audio objects. As explained above, depending on the embodiment the object adapting means may include a separate scene adapter and object refiner as shown in Fig. 1, or may only include one of the scene adapter and object refiner.

Referring now to Fig. 4, a method performed by the object-based audio rendering apparatus 100 of Fig. 1 is illustrated, according to an embodiment of the present invention. First, in step S401 the apparatus 100 receives the audio objects and metadata. Then, in step S402 the scene adapter 101 optimises the sound scene by adapting the objects as necessary, taking into account the current reproduction scenario as explained above. Next, in step S403 the object refiner 102 applies further refinement on an object-by-object basis as necessary and selects an appropriate renderer for each object, again taking into account the factors described above. Then, in step S404 the speaker interface 104 outputs the rendered audio data to the connected speakers to be reproduced.

Embodiments of the present invention have been described in which rendered audio data can be optimised for various listening scenarios, by taking into account factors that will influence the listener's perception of the rendered audio data. This approach provides intelligent rendering apparatuses and methods that can exploit a rich collection of information, and can render to a wider variety of loudspeaker arrangements than is possible with conventional object-based audio rendering techniques. The intelligent rendering apparatuses and methods can determine an appropriate rendering strategy by modifying the sound scene, applying further refinement on an object-by-object basis as necessary, and selecting one or more optimum rendering algorithms for each object.

Furthermore, embodiments of the invention have been described in which one or more microphone signals recorded from the reproduction environment can be analysed to determine the intelligibility of an audio object, such as dialogue. Signal processing techniques for measuring intelligibility are known in the art, and for the sake of brevity a detailed explanation will not be given here. Examples of intelligibility metrics that can be used in embodiments of the present invention include, but are not limited to: binaural distortion-weighted glimpse proportion metric (BiDWGP); distortion-weighted glimpse proportion metric (DWGP); Speech Intelligibility Index (SII) or an adapted binaural version; and the Speech Transmission Index (STI) or an adapted binaural version.

Examples of adaptations that can be applied to audio objects to increase their intelligibility include, but are not limited to: decreasing the relative level of any non-dialogue objects in the sound scene so as to increase the signal-to-noise ratio relative to a dialogue object; changing the spectral balance of the non-dialogue objects to give attenuation in a bandwidth over which speech intelligibility is important, so the dialogue can be more clearly heard; applying effects processing to the non-dialogue objects, for example applying expansion and attenuating these objects to give gaps in the background where the speech can be more easily heard; shifting the spatial location of the non-dialogue and/or dialogue objects so that the listener can more effectively exploit binaural unmasking; changing the spatial rendering of the non-dialogue objects, for example to reduce the correlation of the non-dialogue sounds at the two ears so that the listener can more effectively exploit binaural unmasking; changing the reverberation or early-to-late ratio added to objects; time-shifting the non-dialogue objects so they do not mask the dialogue, or alternatively time-shifting the dialogue to achieve the same effect; and applying speech modification algorithms to the dialogue, the speech modification algorithms being configured to boost intelligibility by reallocating speech energy in a time-frequency domain without increasing speech intensity.

Embodiments of the invention have also been described in which audio objects can be adapted to take into account acoustics of the room or other environment in which the rendered audio will be reproduced. For example, in some embodiments the system configuration information acquiring unit 202 can be configured to acquire information about reverberation, which may be specified by additional metadata which describes the reverberation in terms of early reflections and a late tail. The reflections can be described by various properties, such as direction, timing and equalisation filters, while the tail can be described for each spectral band by properties such as the onset time, attack time, amplitude, and exponential decay time constant. The reverberation metadata can be adapted to modify the overall spatio-temporal response and perceived reverberance in the reproduction room. For example, the decay time of a reverberant object could be reduced so that the reproduced decay time, which includes the effect of the reproduction room, is closer to the production target. It will be understood that this is merely one example, and many other types of adaptation are possible.

The system configuration information may include parameters defining the acoustic response from each loudspeaker to the listener via the room, which can depend on the listener position. This multichannel response can be represented in different ways for use in different types of adaptation. For room equalisation, a complex frequency response can be used. For the adaptation of the time response of reverberation described above, the metadata format for representing reverberant objects may be used. This enables a direct comparison between the response of the room and the incoming reverberant object metadata, without any costly real-time processing. The object based approach can also be advantageous for room equalisation, since only the later part of the reverberant response needs to be filtered such that the overall response in the reproduction room is flat. In contrast, a channel equalisation approach would affect the whole signal including direct sound and early reflections, which is not realistic.

Whilst certain embodiments of the invention have been described herein with reference to the drawings, it will be understood that many variations and modifications will be possible without departing from the scope of the invention as defined in the accompanying claims.

**Claims**

1.      Apparatus for performing object-based audio rendering on a plurality of audio objects which define a sound scene, each audio object comprising at least one audio signal and associated metadata, the apparatus comprising:
        a plurality of renderers each capable of rendering one or more of the audio objects to output rendered audio data; and
        object adapting means for adapting one or more of the plurality of audio objects for a current reproduction scenario, the object adapting means being configured to send the adapted one or more audio objects to one or more of the plurality of renderers.

2.      The apparatus of claim 1, wherein the object adapting means comprises:
        a scene adapter configured to adapt the sound scene for the current reproduction scenario by adapting an audio signal and/or metadata of one or more of the audio objects; and
        an object refiner configured to receive the plurality of audio objects , including any audio objects adapted by the scene adapter, further refine individual ones of the audio objects by adapting an audio signal and/or metadata of said individual audio objects, and send the plurality of audio objects to one or more selected renderers among the plurality of renderers.

3.      The apparatus of claim 2, wherein the object adapting means further comprises:
        a contextual information generator configured to determine one or more parameters relating to the current reproduction scenario, and to control the scene adapter and/or the object refiner based on rules for optimising the rendered audio data according to the determined one or more parameters.

4.      The apparatus of claim 1, 2 or 3, wherein the metadata includes editorial information defining a constraint within which an audio object may be adapted, and the object adapting means is configured to adapt one of the plurality of audio objects within the constraint defined by the metadata.

5.      The apparatus of claim 4, wherein the constraint defined by the editorial information includes a priority order of a plurality of properties of one or more audio objects within the sound scene, and the object adapting means is configured to preferentially modify a property with a lower priority instead of modifying a property with a higher priority, when adapting the one or more audio objects.

6.      The apparatus of any one of the preceding claims, wherein the associated metadata for each of the plurality of audio objects defines an object type, and the object adapting means is configured to be capable of sending different audio objects of the same type to different renderers.

7.      The apparatus of any one of the preceding claims, wherein the object adapting means is configured to adapt one or more of the audio objects according to a user preference setting.

8.      The apparatus of claim 7, wherein the user preference setting includes one or more of:

an intelligibility setting defining a preferred intelligibility level for a dialogue audio object;

an envelopment setting; and

a language setting defining a preferred language.

9.      The apparatus of any one of the preceding claims, further comprising:

an input configured to receive an audio signal from at least one microphone arranged to detect sound generated by the one or more speakers,

wherein the scene adapter is configured to increase intelligibility of a dialogue audio object in response to an increase in a noise level detected by the one or more microphones.

10.      The apparatus of any one of the preceding claims, wherein the object adapting means is configured to determine the current reproduction scenario based on one or more environment parameters relating to a physical environment in which the rendered audio data is to be reproduced.

11.      The apparatus of any one of the preceding claims, wherein the object adapting means is configured to determine the current reproduction scenario based on one or more speaker parameters relating to capabilities of one or more speakers through which the rendered audio data is to be reproduced.

12.      The apparatus of claim 11, further comprising:

a device manager configured to receive information identifying audio output capabilities of one or more connected external devices, and set the one or more speaker parameters according to the received information; and a speaker interface configured to send the rendered audio data to the one or more connected external devices for reproduction.

13. The apparatus of claim 11 or 12, wherein the one or more speaker parameters include one or more of:

speaker number information defining a number of speakers presently connected to the speaker interface;

speaker location and/or orientation information defining a location and/or orientation of at least one of the one or more speakers;

bandwidth information defining a bandwidth of a connection between the speaker interface and one of the one or more speakers;

frequency response information defining a frequency response of at least one of the one or more speakers; and latency information defining a response time of at least one of the one or more speakers.

14. The apparatus of any one of the preceding claims, wherein the object adapting means is configured to determine the current reproduction scenario based on an audience location parameter defining a location and/or orientation of one or more individuals within a space in which the rendered audio data is to be reproduced.

15. An object-based audio rendering method comprising:

receiving a plurality of audio objects which define a sound scene, each audio object comprising at least one audio signal and associated metadata;

adapting one or more of the plurality of audio objects for a current reproduction scenario; and sending the adapted one or more audio objects to one or more of a plurality of renderers to be converted into rendered audio data.

16. The method of claim 15, wherein adapting the one or more audio objects comprises:

adapting the sound scene for the current reproduction scenario by adapting an audio signal and/or metadata of one or more of the audio objects;

further refine individual ones of the audio objects by adapting an audio signal and/or metadata of said individual audio objects; and sending the plurality of audio objects to one or more selected renderers among the plurality of renderers.

17.     A computer-readable storage medium arranged to store computer program instructions which, when executed, perform a method according to claim 15 or 16.

**Abstract**

**Object-Based Audio Rendering**

Apparatus and methods are disclosed for performing object-based audio rendering on a plurality of audio objects which define a sound scene, each audio object comprising at least one audio signal and associated metadata. The apparatus comprises: a plurality of renderers each capable of rendering one or more of the audio objects to output rendered audio data; and object adapting means for adapting one or more of the plurality of audio objects for a current reproduction scenario, the object adapting means being configured to send the adapted one or more audio objects to one or more of the plurality of renderers.

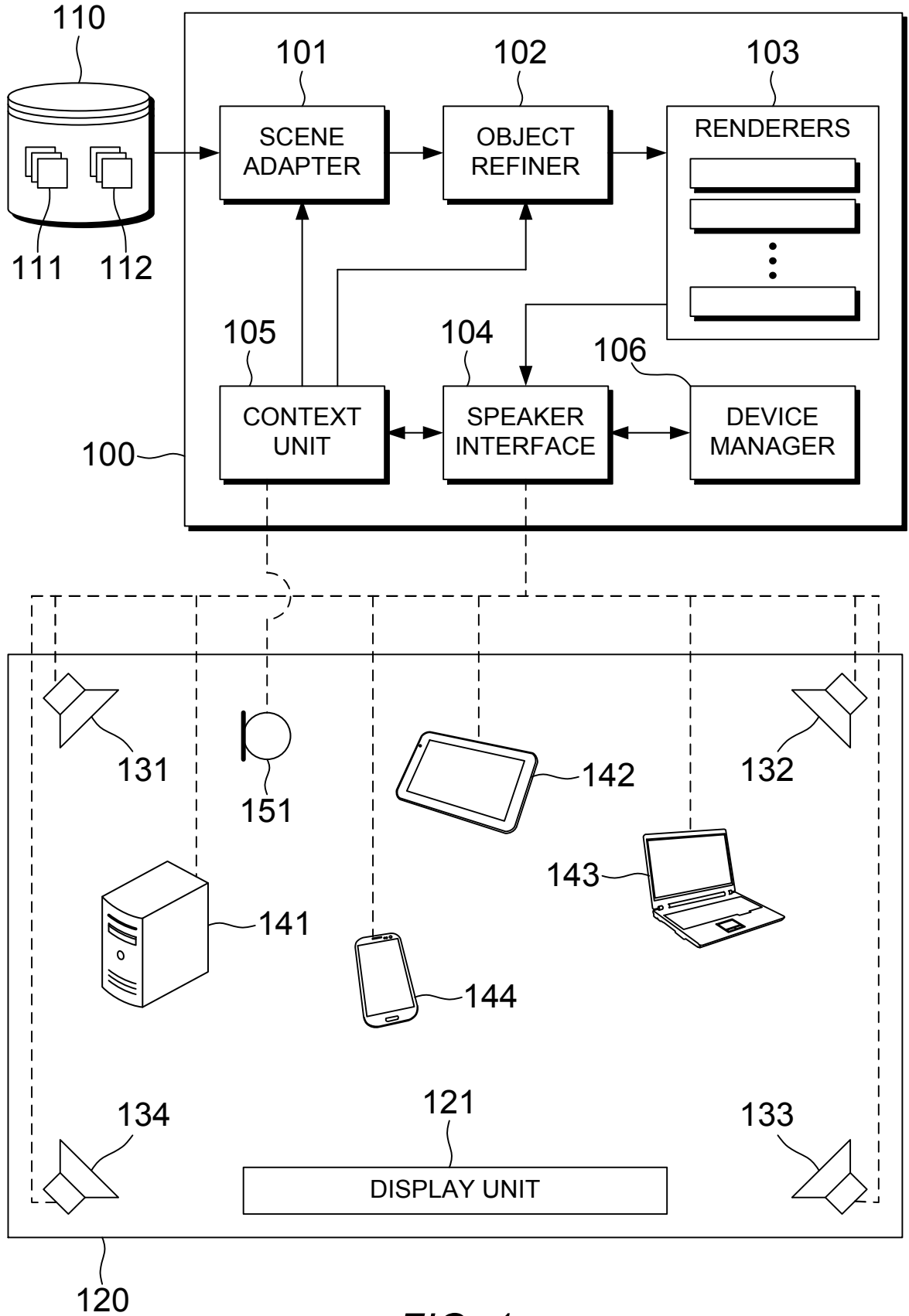

*FIG. 1*

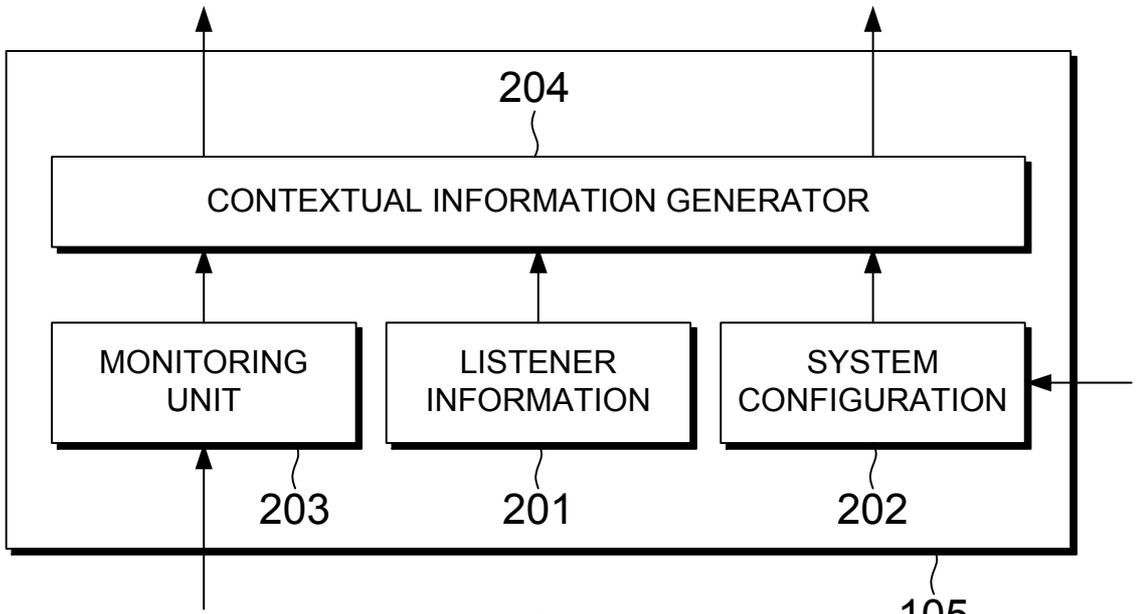

204

CONTEXTUAL INFORMATION GENERATOR

| MONITORING UNIT | LISTENER INFORMATION | SYSTEM CONFIGURATION |
|---|---|---|

203    201    202

*FIG. 2*

105

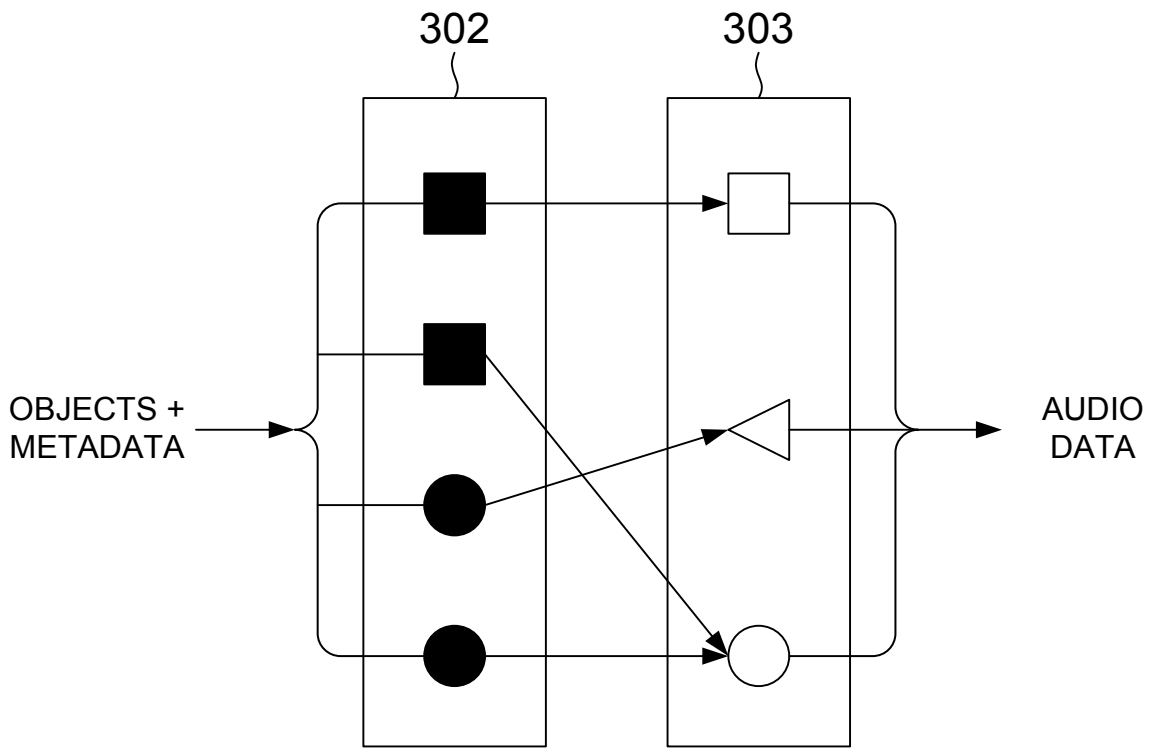

302    303

OBJECTS + METADATA    AUDIO DATA

*FIG. 3*

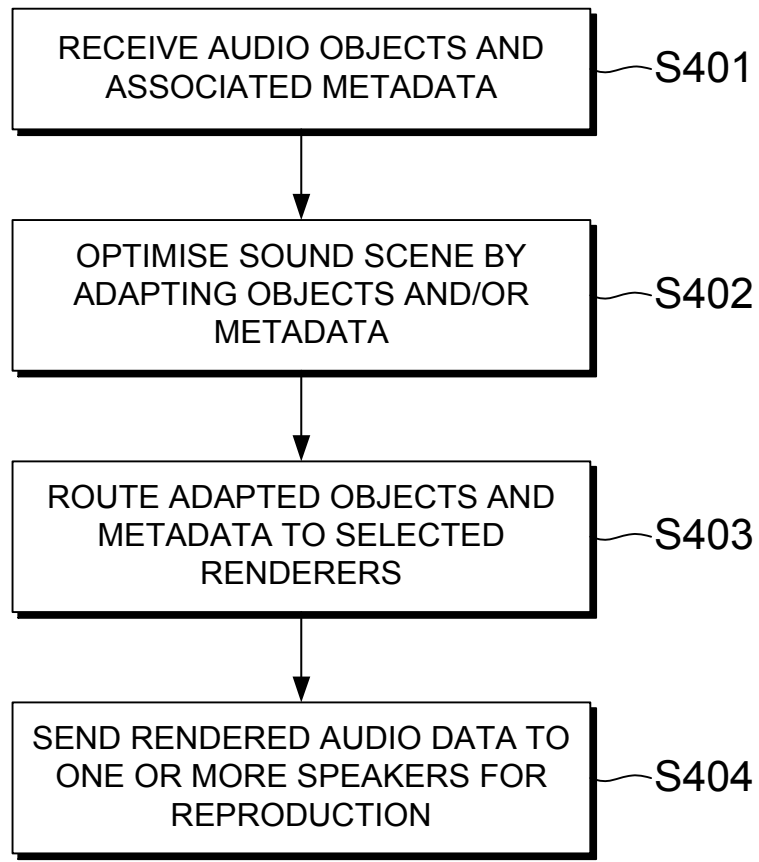

RECEIVE AUDIO OBJECTS AND ASSOCIATED METADATA —S401

OPTIMISE SOUND SCENE BY ADAPTING OBJECTS AND/OR METADATA —S402

ROUTE ADAPTED OBJECTS AND METADATA TO SELECTED RENDERERS —S403

SEND RENDERED AUDIO DATA TO ONE OR MORE SPEAKERS FOR REPRODUCTION —S404

*FIG. 4*

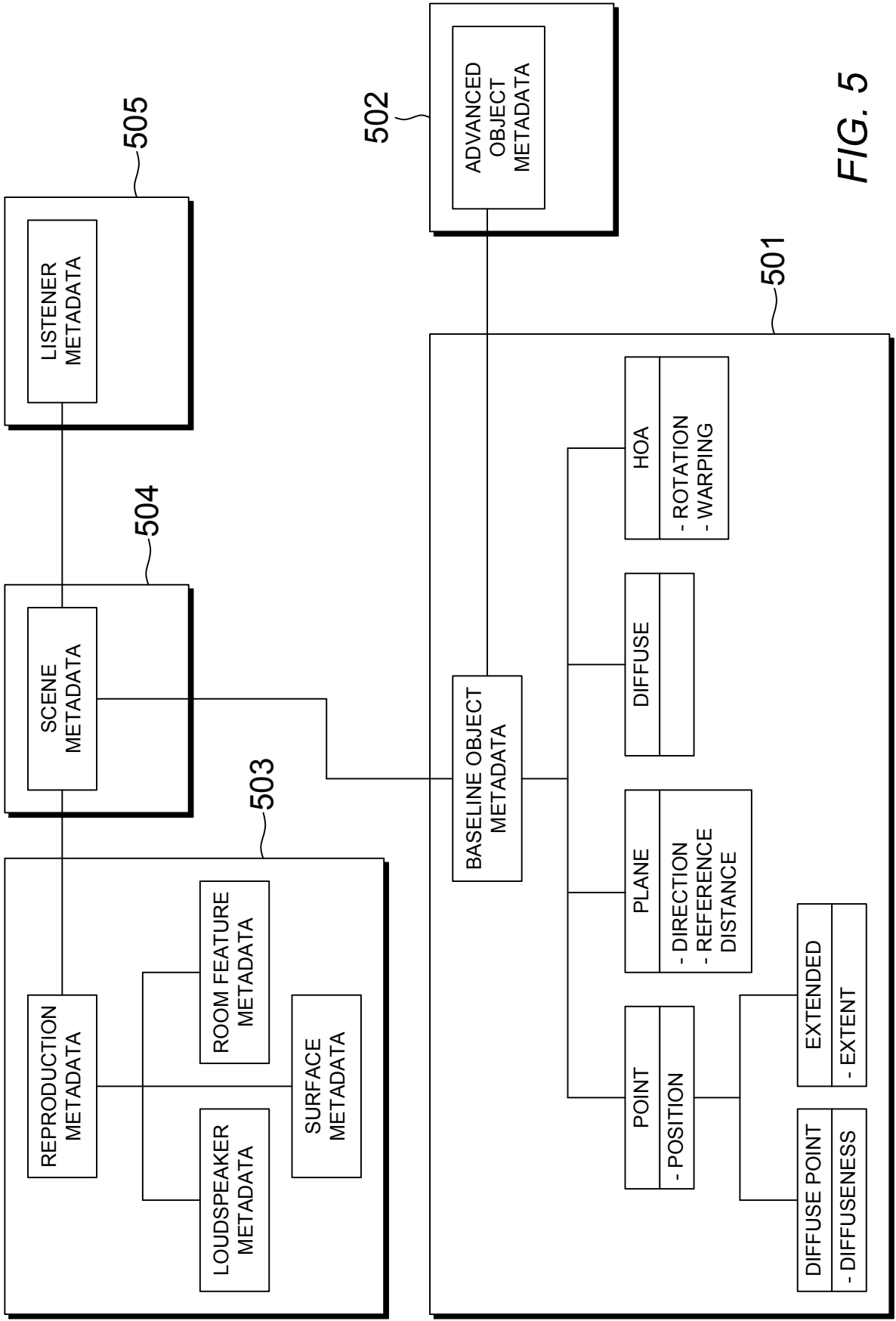

*FIG. 5*